\documentclass[twocolumn,showpacs,amsmath,amssymb]{revtex4}
\usepackage{graphicx}
\usepackage{amsfonts}
\usepackage{dcolumn}
\usepackage{bm}

\usepackage{CJK}

\begin{document}
\title{Large deviation induced phase switch in an inertial majority-vote model}

\author{Hanshuang Chen$^{1}$}\email{chenhshf@ahu.edu.cn}

\author{Chuansheng Shen$^{2,3}$}\email{csshen@mail.ustc.edu.cn}

\author{Haifeng Zhang$^4$}

\author{J\"urgen Kurths$^{3,5}$}\email{Juergen.Kurths@pik-potsdam.de}

\affiliation{$^{1}$School of Physics and Materials Science, Anhui University, Hefei, 230601, China \\
$^2$Department of Physics, Anqing Normal
University, Anqing, 246011, China\\
$^3$Department of Physics, Humboldt University, 12489 Berlin, Germany \\
$^4$School of Mathematical Science, Anhui University, Hefei, 230601, China \\
\ $^5$Potsdam Institute for Climate Impact Research (PIK), 14473
Potsdam, Germany }

\date{\today}

\begin{abstract}
We theoretically study noise-induced phase switch phenomena in an
inertial majority-vote (IMV) model introduced in a recent paper
[Phys. Rev. E 95, 042304 (2017)]. The IMV model generates a strong
hysteresis behavior as the noise intensity $f$ goes forward and
backward, a main characteristic of a first-order phase transition,
in contrast to a second-order phase transition in the original MV
model. Using the Wentzel-Kramers-Brillouin approximation for the
master equation, we reduce the problem to finding the zero-energy
trajectories in an effective Hamiltonian system, and the mean
switching time depends exponentially on the associated action and
the number of particles $N$. Within the hysteresis region, we find
that the actions along the optimal forward switching path from
ordered phase (OP) to disordered phase (DP) and its backward path,
show distinct variation trends with $f$, and intersect at $f=f_c$
that determines the coexisting line of OP and DP. This results in a
nonmonotonic dependence of the mean switching time between two
symmetric OPs on $f$, with a minimum at $f_c$ for sufficiently large
$N$. Finally, the theoretical results are validated by Monte Carlo
simulations.

\end{abstract}
 \maketitle

\textbf{Noise induced phase switch between coexisting stable phases
underlies many important physical, chemical, biological, and social
phenomena. Examples include diffusion in solids, switching in
nanomagnets and Josephson junctions, nucleation, chemical reactions,
protein folding, and epidemics. In this paper, we apply the
Wentzel-Kramers-Brillouin approximation to study noise-induced phase
switch phenomena in an inertial majority-vote model. The mean
switching time is determined by the classical action along the
zero-energy trajectories in an effective Hamiltonian system. The
results show that the mean switching time between two symmetric
ordered phases depends nonmonotonically on the noise intensity. This
attributes to the first-order characteristic of phase transition of
the model leading to occurrence of a stable disordered phase within
the hysteresis region. Our results shed some new understanding for
the nontrivial role of noise in social systems.}

\section{Introduction}
Spin models like the Ising model play a fundamental role in studying
phase transitions and critical phenomena in the field of statistical
physics and many other disciplines \cite{RMP08001275}. They have
also significant implications for understanding social phenomena
where co-ordination dynamics is observed, e.g. in consensus
formation and adoption of innovations \cite{RMP09000591}. The spin
orientations can represent the choices made by an agent on the basis
of information about its local neighborhood.

One of the simplest nonequilibrium generalizations of the Ising
model, called the majority-vote (MV) model, was proposed by Oliveira
in 1992 \cite{JSP1992}. The model displays an up-down symmetry and a
continuous second-order order-disorder phase transition at a
critical value of noise. Studies on regular lattices showed that the
critical exponents are the same as those of the Ising model
\cite{JSP1992,PhysRevE.75.061110,PhysRevE.81.011133,PhysRevE.89.052109,PhysRevE.86.041123},
in accordance with the conjecture by Grinstein \emph{et al.}
\cite{PhysRevLett.55.2527}. The MV model has also been extensively
studied for various interacting substrates, including random graphs
\cite{PhysRevE.71.016123,PA2008}, small world networks
\cite{PhysRevE.67.026104,IJMPC2007,PA2015}, scale-free networks
\cite{IJMPC2006(1),IJMPC2006(2),PhysRevE.91.022816}, and some others
\cite{PhysRevE.77.051122,PA2011}. These studies have shown that the
universality classes and the critical exponents depend on the
topologies of the underlying interacting substrates.

In a recent paper \cite{PhysRevE.95.042304}, we have incorporated an
inertial effect into the microscopic dynamics of the spin flipping
of the MV model, where the spin-flip probability of any individual
depends not only on the states of its neighbors, but also on its own
state. In contrast to a continuous second-order phase transition in
the original MV model, the inertial MV (IMV) model generates a
discontinuous first-order phase transition. Such a discontinuous
phase transition is manifested by a strong hysteresis behavior as
the noise intensity goes forward and backward. Within the hysteresis
region, the stochastic fluctuations can induce switches to occur
between ordered phase (OP) and disordered phase (DP).

In the present work, we aim to study the switching phenomena based
on the Wentzel-Kramers-Brillouin (WKB) approximation for the master
equation
\cite{PhysRevE.70.041106,JSP127.861,PhysRevLett.97.200602,PhysRevE.75.031122,PhysRevLett.101.078101,PhysRevLett.103.068101,PhysRevLett.101.268103,PhysRevE.81.021116}.
This method has been used to study large deviation-induced
phenomena, e.g. extinction process in a system with an absorbing
state (See \cite{WKBReview1,WKBReview2} for two recent reviews). The
WKB approximation converts the master equation governing the
stochastic spin-flipping processes of the IMV model into an
effective Hamiltonian system. This enables us to calculate the mean
switching time between different phases given by the classic action
along the optimal switching path (zero-energy trajectory). Within
the hysteresis region, the coexisting line is determined when the
actions along the forward and backward switching path between OP and
DP meet, where OP and DP are of equivalent stability. Interestingly,
due to the existence of the stable DP, the mean switching time
between two OPs shows a nonmonotonic dependence on the noise
intensity.

\section{Model}
We consider a system of $N$ spins, where each spin $i$ is denoted by
either $\sigma_i=+1$ (up) or $\sigma_i=-1$ (down). In each time
step, we first randomly choose a spin $i$ and then randomly choose
$q$ other spins as its neighborhood, labeled by $i_1, \cdots, i_q$,
where $q$ is the number of neighbors. We then try to flip the spin
$i$ with the probability,
\begin{eqnarray}
w_i(\sigma)=\frac{1}{2}\left[{1-(1-2f)\sigma_i S\left( {\Theta
_i}\right)}\right], \label{eq1}
\end{eqnarray}
where
\begin{eqnarray}
\Theta_i =(1-\theta)\sum\limits_{j=1}^q \sigma_{i_j} /q + \theta
\sigma_i\label{eq2}
\end{eqnarray}
is the local field of spin $i$. $S(x)=sgn(x)$ if $x\neq0$ and
$S(0)=0$, $f\in[0,\kern 2pt 0.5]$ is a noise parameter, and
$\theta\in[0, \kern 2pt 0.5)$ is a parameter controlling the weight
of the inertia. The larger the value of $\theta$, the greater is the
inertia of the system. For $\theta=0$, we recover the original MV
model without inertial effect. Since the value of $q$ does not
change qualitatively the results in the present work, $q=20$ is
fixed throughout the paper.

The MV model does not only play an important role in the study of
nonequilibrium phase transitions, but also helps to understand
opinion dynamics in social or biological systems. In this model,
binary spins can represent two opposite opinions, or competitive
language features, and the noise parameter $f$ plays the role of the
temperature in equilibrium systems and measures the probability of
aligning antiparallel to the majority of neighbors. Moreover,
consideration of the inertial effect is based on the fact that
individuals in a social or biological context have a tendency for
beliefs to endure once formed. In a recent experiment
\cite{NatPhys2014(2)}, behavioral inertia was found to be essential
for collective turning of starling flocks. A counterintuitive
``slower is faster" effect of the inertia on ordering dynamics of
the voter model was shown in \cite{PhysRevLett.101.018701}.

\section{Mean-field theory}
To begin, we proceed our analysis from the deterministic mean-field
theory. Let $x=n/N$ denote the density of up spins, where $n$ is the
number of up spins. The rate equation for the density $x$ can be
written as
\begin{eqnarray}
\dot x =-x w_+(x)+(1-x) w_-(x),\label{eq3}
\end{eqnarray}
where $w_+(x)$ and $w_-(x)$ are the flipping probabilities of an up
spin and a down spin, respectively. According to Eq.(\ref{eq1}),
$w_+(x)$ can be written as the sum of three parts,
\begin{eqnarray}
w_+(x)=f P_>^+ + \frac{1}{2} P_=^+ + (1-f) P_<^+,\label{eq4}
\end{eqnarray}
where $P_>^+$ ($P_=^+$ and $P_<^+$) is the probability that the
local field of an up spin is larger than zero (equals to zero and is
less than zero). Likewise, $w_-(x)$ can be written as
\begin{eqnarray}
w_-(x)=(1-f) P_>^- + \frac{1}{2} P_=^- + f P_<^-,\label{eq5}
\end{eqnarray}
where $P_>^-$ ($P_=^-$ and $P_<^-$) is the probability that the
local field of a down spin is larger than zero (equals to zero and
is less than zero). These probabilities can be expressed by binomial
distribution
\begin{eqnarray}
  P_>^\pm &=&\sum\limits_{k={\left\lceil q_\pm \right\rceil}}^q  C_q^k x^k (1-x)^{q - k}, \label{eq6} \\
  P_=^\pm &=&{\delta _{\left\lceil {{q_ \pm }} \right\rceil ,{q_ \pm }}}C_q^{\left\lceil {{q_ \pm }} \right\rceil }{x^{\left\lceil {{q_ \pm }} \right\rceil }}{(1 - x)^{q - \left\lceil {{q_ \pm }} \right\rceil
  }}, \label{eq7} \\
  P_<^\pm &=&\sum\limits_{k=0}^{\left\lfloor q_\pm \right\rfloor} C_q^k x^k (1-x)^{q -
  k},\label{eq8}
\end{eqnarray}
where $\left\lceil \cdot \right\rceil$ ($\left\lfloor \cdot
\right\rfloor$) is the ceiling (floor) function, $\delta$ is the
Kronecker symbol, and $C_q^ k = q!/[k!(q- k)!]$ are the binomial
coefficients. $q_+  = (1-2\theta)q/[2(1- \theta)]$ and $q_- =
q/[2(1- \theta)]$ are the number of up-spin neighbors of an up spin
and a down spin satisfying $\Theta = 0$, respectively. It is clear
that $q_+ + q_-=q$ holds for any $\theta$ and
${P_>^\pm}+{P_=^\pm}+{P_<^\pm}=1$ due to probability conservation.

\begin{figure}
\centerline{\includegraphics*[width=0.8\columnwidth]{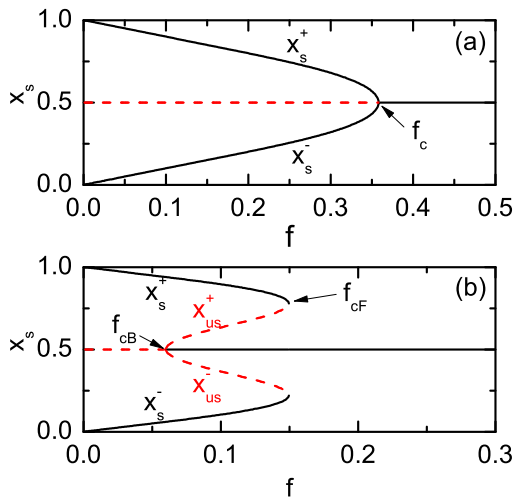}}
\caption{The stationary solution $x_s$ of Eq.(\ref{eq3}) as a
function of noise parameter $f$ for $\theta=0$ (a) and for
$\theta=0.35$ (b). \label{fig1}}
\end{figure}

\begin{figure}
\centerline{\includegraphics*[width=0.8\columnwidth]{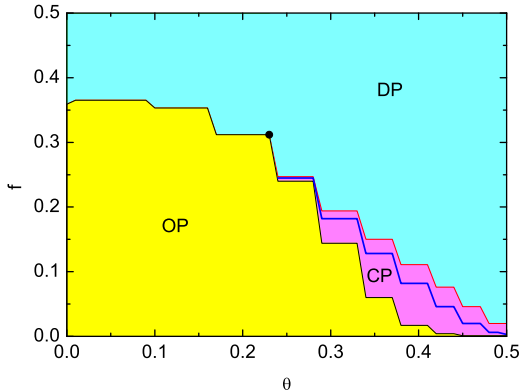}}
\caption{Phase diagram in the $(\theta, f)$ plane. The phase diagram
is divided into three regions: OP, CP, and DP. They collide at a
triple point (solid circle). The blue line indicates the coexisting
line of OP and DP. Within the CP region, OP is stable and DP is
metastable below the coexisting line, but DP is stable and OP is
metastable above the coexisting line. \label{fig2}}
\end{figure}

Since $w_+=w_-$ at $x = 1/2$, one can easily check that $x = 1/2$ is
always a stationary solution of Eq.(\ref{eq3}). This trivial
solution corresponds to a DP. The other possible solutions can be
obtained by numerically solving Eq.(\ref{eq3}). For $\theta=0$, it
is well-known that the standard MV model undergoes a continuous
second-order phase transition from an OP (there exist two symmetric
stable solutions of $x_s^{\pm}\neq 1/2$) to a DP at $f=f_c$, as
shown in Fig.\ref{fig1}(a), where the critical noise $f_c$ is
determined by which the trivial solution $x=1/2$ loses its
stability. While for a large enough $\theta$, e.g. for $\theta=0.35$
as shown in Fig.\ref{fig1}(b), the IMV model undergoes a
discontinuous first-order phase transition. A hysteresis loop occurs
in the range $f_{c_B}<f<f_{c_F}$. In detail, for $f<f_{c_B}$ the
model has two symmetric stable solutions $x_s^{\pm}$ and an unstable
solution $x=1/2$, and thus OP is stable in the region. For
$f>f_{c_F}$, the DP is the only stable phase. Within the hysteresis
region ($f_{c_B}<f<f_{c_F}$), the model is bistable with the
coexisting phase (CP) of OP and DP. In this case, there exist three
stable solutions. Two of them are symmetric stable solutions
$x_s^{\pm}$, and the other one is $x=1/2$. Between $x_s^+$ ($x_s^-$)
and $x=1/2$, there is an unstable solution $x_{us}^+$ ($x_{us}^-$).
The model evolves into either OP or DP depending on the initial
value of $x$.

Figure \ref{fig2} shows the global phase diagram in the $(\theta,
f)$ plane. It is separated into three regions by $f_{c_B}$ and
$f_{c_F}$: OP, CP, and DP. These three regions collide at a
so-called tricritical point, $(\theta^*, f^*)=(0.23, 0.312)$. The
phase transition is of second order if $\theta<\theta^*=0.23$ and of
first order if $\theta>\theta^*$.

\begin{figure}
\centerline{\includegraphics*[width=0.8\columnwidth]{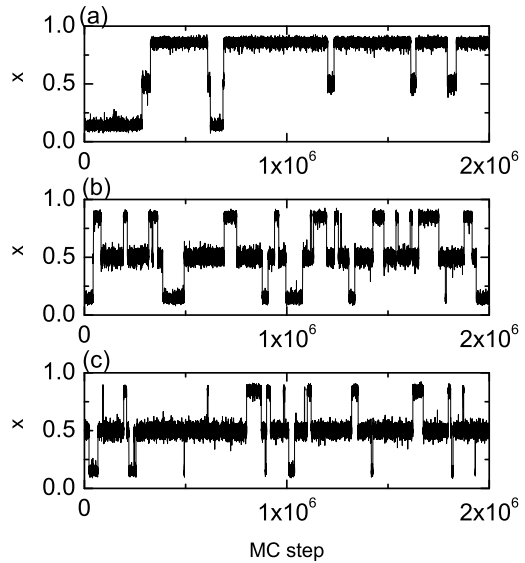}}
\caption{Phase switches. Time series of the density $x$ of up spins
for three different noises: $f=0.124$ (a), $0.127$ (b), and $0.129$
(c). The other parameters are $N=400$ and $\theta=0.35$.
\label{fig3}}
\end{figure}

\section{Master equation and WKB theory}
However, for a finite size system stochastic fluctuations can induce
switches from one phase to another one. In Fig.\ref{fig3}, we show
three typical time series of $x$ within the coexisting region for
$N=400$ and $\theta=0.35$ obtained from Monte Carlo (MC)
simulations, where one MC step is defined as each spin is attempted
to flip once on average. It can be observed that the switching
phenomena between OP and DP rarely occur. As $f$ increases, it seems
that the switch from OP to DP happens more frequently, but its
backward switch happens more infrequently. Therefore, an interesting
question arises: how does one calculate the mean switching time from
a theoretical perspective? Since the mean-field treatment ignores
the effect of stochastic fluctuations, it fails to account for
switching phenomena induced by large deviations. To this end, let
$P_n(t)$ be the probability that the number of up spins is $n$ at
time $t$. The master equation for $P_n(t)$ reads,
\begin{eqnarray}
\frac{{d{P_n}(t)}}{{dt}} = && {W_+}(n+1)P_{n+1}(t) + {W_-}(n-1)P_{n
- 1}(t) \nonumber \\ &&-\left[ {W_+}(n)+{W_-}(n) \right]{P_n}(t).
\label{eq9}
\end{eqnarray}
Here, $W_+(n)$ and $W_-(n)$ are the respective rates of flipping up
spins and down spins, which can be written as
\begin{eqnarray}
  {W_ + }(n) &=& n{w_ + }(x), \label{eq10}\\
  {W_ - }(n) &=& (N - n){w_-}(x). \label{eq11}
\end{eqnarray}

By employing the WKB approximation for the probability $P_n$
\cite{WKBReview1}, we write
\begin{eqnarray}
 P_n=e^{N\mathcal {S}(x)}. \label{eq12}
\end{eqnarray}
As usual, we assume that $N$ is large and take the leading order in
a $N^{-1}$ expansion, by writing $P(n \pm 1)\approx P_n e^{\mp
\partial S/\partial x}$ and $W(n \pm 1)\approx W(n)$. We then arrive at
the Hamilton-Jacobi equation
\begin{eqnarray}
\frac{{\partial \mathcal {S}}}{{\partial t}} + \mathcal {H}(x,p) =
0, \label{eq13}
\end{eqnarray}
where $\mathcal {S}$ and $\mathcal {H}$ are called the action and
Hamiltonian, respectively. As in classical mechanics, $\mathcal {H}$
is a function of the coordinate $x$ and its conjugate momentum $p
=\partial S/\partial x$:
\begin{eqnarray}
\mathcal {H}(x,p)= {\bar w}_-(x)(e^p- 1)+ {\bar w}_+(x)(e^{-p}-1),
\label{eq14}
\end{eqnarray}
where $\bar w_{\pm}(x)=W_{\pm}(n)/N$ are the rescaled rates. We then
write the canonical equations of motion,
\begin{eqnarray}
  \dot x &=& {\partial_p}\mathcal {H}\left( {x,p} \right) = {{\bar w}_-}(x){e^p} - {{\bar w}_ + }(x){e^{ - p}}, \label{eq15} \\
  \dot p &=&  - {\partial_x}\mathcal {H}\left( {x,p} \right) = {{\bar w'}_ - }(x)\left( {1 - {e^p}} \right) + {{\bar w'}_ + }(x)\left( {1 - {e^{ - p}}}
  \right), \nonumber \\ \label{eq16}
\end{eqnarray}
where $\bar w'_ +$ and $\bar w'_ -$ are the derivative of $\bar w_
+$ and $\bar w_ -$ with regard to $x$, respectively. We focus on the
switching trajectory from one phase to another one. This implies
that there will be some trajectory along which $\mathcal {S}$ is
minimized, which represents the maximal probability of such a
switching event. This corresponds to the zero-energy ($\mathcal {H}
= 0$) trajectory in the phase space $(x,p)$ from one fixed point to
another one. According to equation (14), $\mathcal {H} = 0$ implies
that
\begin{eqnarray}
p = 0 {\kern 10pt}  {or} {\kern 10pt} p=\ln {\frac{\bar w_+(x)}{\bar
w_-(x)}}. \label{eq17}
\end{eqnarray}
In particular, the line $p = 0$ corresponds to the result of the
mean-field theory, as Eq.(\ref{eq15}) for $p=0$ recovers to the
mean-field equation (\ref{eq3}).

\begin{figure}
\centerline{\includegraphics*[width=1.0\columnwidth]{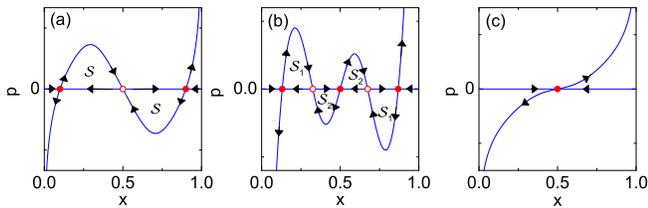}}
\caption{Zero-energy trajectories in the $(x, p)$ phase space for
three different regions: OP (a), CP (b), and DP (c). All fixed
points (circles) correspond to the stationary solutions of the
mean-filed equation (\ref{eq3}), including stable (solid circles)
and unstable (empty circles) solutions. \label{fig4}}
\end{figure}

\begin{figure}
\centerline{\includegraphics*[width=0.8\columnwidth]{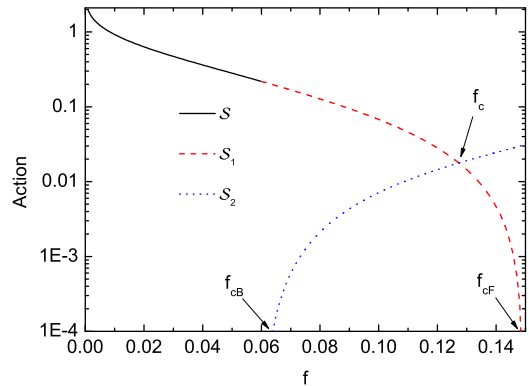}}
\caption{The actions as a function of $f$ for $\theta=0.35$.
$\mathcal {S}_1$ and $\mathcal {S}_2$ meet at $f=f_c$ where the OP
and DP are equivalently stable.  \label{fig5}}
\end{figure}

Figure \ref{fig4} depicts three typical zero-energy curves in the
$(x, p)$ plane, corresponding to OP (Fig.\ref{fig4}(a)), CP
(Fig.\ref{fig4}(b)), and DP (Fig.\ref{fig4}(c)), respectively. These
curves determine the topologies of the phase space. All the stable
and unstable solutions of $x$ in the mean-field theory become saddle
points in phase space. When the model is in OP (Fig.\ref{fig4}(a)),
the mean switching time from one OP to another OP is determined by
$\left\langle T \right\rangle  \sim {e^{N\mathcal {S}}}$, where
$\mathcal{S}= \int_{x_s^\pm}^{1/2}{{\ln {\frac{\bar w_+(x)}{\bar
w_-(x)}}}} dx$ is the action from $x_s^-$ to $x_s^+$ or from $x_s^+$
to $x_s^-$ along the zero-energy trajectory. When the model is in CP
(Fig.\ref{fig4}(b)), the mean switching time from one OP to DP is
determined by $\left\langle T_1 \right\rangle \sim {e^{N\mathcal
{S}_1}}$, and the mean switching time from DP to one of the OPs is
determined by $\left\langle T_2 \right\rangle \sim {e^{N\mathcal
{S}_2}}$, where $\mathcal {S}_1 = \int_{{x_{s}^ \pm }}^{x_{us}^\pm}
{\ln \frac{{{{\bar w}_ + }(x)}}{{{{\bar w}_ - }(x)}}} dx$ is the
action from $x_s^-$ ($x_s^+$) to $x=1/2$ along the zero-energy
trajectory, and $\mathcal {S}_2 = \int_{{x_{us}^ \pm }}^{1/2} {\ln
\frac{{{{\bar w}_ + }(x)}}{{{{\bar w}_ - }(x)}}} dx$ is the action
from $x=1/2$ to $x_{s}^-$ ($x_{s}^+$) along the backward trajectory.
Thus, the mean switching time from one OP to another OP is
$\left\langle T \right\rangle=\left\langle T_1
\right\rangle+\left\langle T_2 \right\rangle$. When the model is in
DP (Fig.\ref{fig4}(c)), the only stable phase is disordered and
there is no switching phenomenon.

Figure \ref{fig5} shows the actions as a function of $f$ for
$\theta=0.35$. In the OP region, $\mathcal {S}$ decreases
monotonically with $f$. In the CP region, $\mathcal {S}_1$ and
$\mathcal {S}_2$ exhibit distinct variations with $f$, and they
intersect at $f=f_c$ where the OP and DP have the same stability.
This implies that for $f_{c_B}<f<f_c$ the OP is more stable and for
$f_c<f<f_{c_F}$ the OP is less stable than the DP. We have
calculated $f_c$ as a function of $\theta$ (coexisting line), as
shown by the blue line in Fig.\ref{fig2}. Since the mean switching
times from one OP to DP and then to another OP are both
exponentially dependent on the corresponding actions and the number
of spins, the mean switching time $\left\langle T \right\rangle$
between the two ordered phases is dominated by the larger one of
$\mathcal {S}_1$ and $\mathcal {S}_2$. Therefore, one can expect
that $\left\langle T \right\rangle$ will change non-monotonically
with $f$. A minimum in $\left\langle T \right\rangle$ will locate at
$f=f_c$ for enough large $N$.

\begin{figure}
\centerline{\includegraphics*[width=0.8\columnwidth]{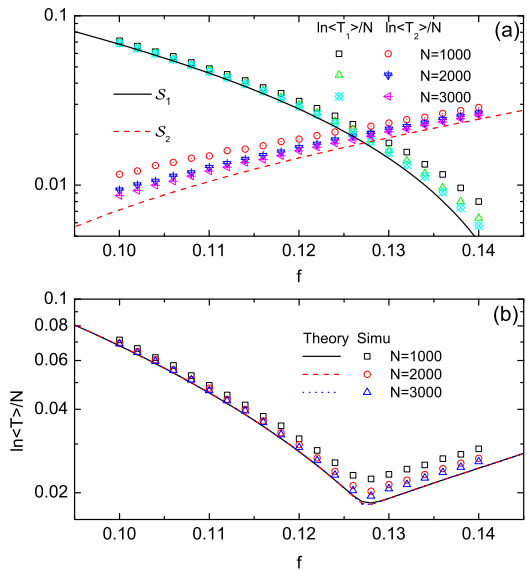}}
\caption{(a) $\ln\left\langle T_1 \right\rangle/N$ and
$\ln\left\langle T_2 \right\rangle/N$ as a function of $f$ for
different $N$ at $\theta=0.35$. The lines indicate the actions
$\mathcal {S}_1$ and $\mathcal {S}_2$. (b) $\ln\left\langle T
\right\rangle/N$ as a function of $f$ for different $N$ at
$\theta=0.35$. The lines indicate the theoretical prediction.
\label{fig6}}
\end{figure}

\section{Numerical Validation}
To obtain the mean switching time, one needs to perform long-time MC
simulations to sample enough switching events. However, the phase
switch is a rare event that occurs very infrequently, especially for
large $N$. Thus, the conventional brute-force simulation becomes
prohibitively inefficient. To overcome this difficulty, we employ a
rare-event sampling method, the forward flux sampling (FFS)
\cite{PRL05018104,JPH09463102}, combined with MC simulation. The FFS
method first defines an order parameter to distinguish between the
initial phase $\mathcal{I}$ and the final phase $\mathcal{F}$, and
then uses a series of interfaces to force the system from
$\mathcal{I}$ to $\mathcal{F}$ in a ratchet-like manner. Here, it is
convenient to select the number of up spins $n$ as the order
parameter. A series of non-intersecting interfaces $n_i$
($0<i<N_{in}$) lie between states $\mathcal{I}$ and $\mathcal{F}$,
such that any path from $\mathcal{I}$ to $\mathcal{F}$ must cross
each interface without reaching $n_{i+1}$ before $n_i$. The
algorithm first runs a long-time simulation which gives an estimate
of the flux $\Phi_{\mathcal{I},0}$ escaping from the basin of
$\mathcal{I}$ and generates a collection of configurations
corresponding to crossings of interface $n_0$. The next step is to
choose a configuration from this collection at random and to use it
to initiate a trial run which is continued until it either reaches
$n_1$ or returns to $n_0$. If $n_1$ is reached, we store the
configuration of the end point of the trial run. We repeat this
step, each time choosing a random starting configuration from the
collection at $n_0$. The fraction of successful trial runs gives an
estimate of of the probability of reaching $n_1$ without going back
into $\mathcal{I}$, $P\left( {n_1 |n_0} \right)$. This process is
repeated, step by step, until $n_{N_{in}}$ is reached, giving the
probabilities $P\left( {n_{i+1} |n_i} \right)$ ($i=1, \cdots,
N_{in}-1$). Finally, we obtain the mean switching time from state
$\mathcal {I}$ to state $\mathcal {F}$,
\begin{equation}
\left\langle T \right\rangle  = \frac{1}{\Phi_{\mathcal{I},0}
\prod\nolimits_{i=0}^{N_{in}-1}{P\left( {n_{i + 1} |n_i}
\right)}}.\label{eq18}
\end{equation}

Fig.\ref{fig6}(a) shows $\ln\left\langle T_1 \right\rangle/N$ and
$\ln\left\langle T_2 \right\rangle/N$ as a function of $f$ for
different $N$ at $\theta=0.35$, where the unit of the mean switching
time is MC step. For comparison, we also show $\mathcal {S}_1$ and
$\mathcal {S}_2$ as a function of $f$, as shown by lines in
Fig.6(a). As predicted by our theoretical analysis, $\ln\left\langle
T_1 \right\rangle/N$ and $\ln\left\langle T_2 \right\rangle/N$ meet
at $f=f_c$. When the mean switching time is relatively small, e.g.,
$\left\langle T_1 \right\rangle$ for large $f$ and $\left\langle T_2
\right\rangle$ for low $f$, there are quantitative discrepancies
between simulation and theory. This is because in this case the
pre-exponential factor of the mean switching time is comparable to
its exponential contribution, but it was not considered in our
analysis. Fig.\ref{fig6}(b) shows $\ln\left\langle T
\right\rangle/N$ as a function of $f$. One can see that there exists
a minimal $\left\langle T \right\rangle$ at $f=f_c$. This implies
that within the CP region an increase in noise from $f_c$ can lead
to a decreases in the mean switching time between two OPs. This
counterintuitive effect is attributed to the appearance of the
stable DP between two OPs, and thus the switch from one OP to
another OP becomes a two-step process.

\section{Conclusions}
In the present work, we have studied phase switch phenomena induced
by large fluctuations in an inertial MV model with a first-order
order-disorder phase transition. By employing the WKB approximation
for the master equation, the mean switching time was evaluated by
the classical action along the optimal switching path in an
effective Hamiltonian system. Within the hysteresis region, the mean
switching time from OP to DP and that from DP to OP show opposite
variation trends with noise $f$, which leads to a minimal mean
switching time between two OPs occurring at the coexisting line
$f_c$ where OP and DP are of the same stability. This implies that
for $f_c<f<f_{c_F}$ noise is unfavorable for the switches between
two OPs. As discussed before, behavioral inertia and noise are both
essential for social dynamics, and therefore our work may provide a
new understanding for their interplay in switch phenomena of social
systems, such as the emergence of a consensus and decision making
\cite{PhysRevLett.94.178701,PhysRevLett.116.038701}, as well as the
spontaneous formation of a common language or culture
\cite{RMP09000591,PhysRevLett.85.3536}. Since the interacting
structures among agents are also vital for dynamics on them, it
would be worth studying the phases switches on a networked inertial
MV model \cite{EPL108.58008,PhysRevLett.117.028302}.

\begin{acknowledgments}
We acknowledge the supports from the National Natural Science
Foundation of China (Grants  No. 11475003, No. 61473001, No.
11205002), the Key Scientific Research Fund of Anhui Provincial
Education Department (Grants No. KJ2016A015) and ``211" Project of
Anhui University (Grants No. J01005106).
\end{acknowledgments}


\end{document}